\def\beg{\begin{equation}}
\def\eeq{\end{equation}}
\documentstyle[12pt]{article}
\begin{document}
\begin{center}
{\Large{\bf Comments on ``Gauge Theory of Composite Fermions: Particle-Flux Separation in Quantum Hall Systems" by I. Ichinose and Tetsuo Matsui, cond-mat/0210142, v 2, 7 Feb.2003 }}
\vskip0.35cm
{\bf Keshav N. Shrivastava}
\vskip0.25cm
{\it School of Physics, University of Hyderabad,\\
Hyderabad  500046, India}
\end{center}

We find that work of Ichinose requires far too many quasiparticles. 
As a result, too many parameters are introduced to fit the mass of a composite fermion (CF) and hence experimentally, it will not be possible to identify all of such quasiparticles and their masses. Further, according to Ichinose, an electron decay should occur but it has not been found. It is much too unrealistic to expect an electron decay similar to the neutron decay. If the CF is really found it is not going to be relevent to the quantum Hall effect data. Therefore, the composite fermion (CF) model of quantum Hall effect is not well founded and should be discarded.

\vfill
Corresponding author: keshav@mailaps.org\\
Fax: +91-402-301 0145.Phone: 301 0811.
\newpage
\baselineskip22pt
\noindent {\bf 1.~ Introduction}

     It has been suggested by Ichinose et al[1] that there is a particle-flux separation (PFS) applicable to electrons. The electron splits into a $``chargeon"$ which carries the charge and a $``fluxon" $ which carries the magnetic field. The $``chargeons"$ are similar to composite fermions (CFs) which have been utilized to understand the quantum Hall effect data in GaAs/AlGaAs. It has been suggested that PFS is similar to charge-spin separation (CSS) in which electron dissociates into holon and spinon with holon carrying the charge but not the spin and the spinon carrying spin but not the charge. There is a concept of phase transition so that there is a critical temperature at which the particle-flux separation occurs. The electron splits into two particles, a $\eta_x$, $``chargeon"$ which is a fermion and a $``fluxon"$ which is a boson,
\beg
electron = chargeon(fermion, \eta_x) + fluxon (boson,\phi_x).
\eeq
The masses of these particles are $m$ for the electron mass in vacuum, $m_e$ for the electron mass in a solid, $m_{\phi}$ the fluxon mass and $m_\eta$ the chargeon mass. The fermion operator $\psi_x$ is given by the product of a fermion operator and a boson operator,
\beg
\psi_x = \phi_x\eta_x.
\eeq
Actually the correct algebra in such a case would require a product wave function which has products of fermion wave function, chargeon wave function and the fluxon wave function. Then there are creation and annihilation operators of the three quasiparticles as,
\beg
\Psi = \psi_x\phi_x\eta_x
\eeq
The number densities are given by the usual product, e.g., 
\beg
\phi_x^\dagger\phi_x= n.
\eeq
The conclusion of this study is that (a) there are quasiparticles, electrons in vacuum, electrons in a solid, chargeons and fluxons. By attaching two fluxons to one electron, we make a composite fermion (CF). In addition to these there are spinons and holons. The electron splits into two quasiparticles, the spinon carries the spin but not the charge and the holon which carries the charge but not the spin,
\beg
electron = spinon + holon
\eeq

The bound state of a $``chargeon"$ and a $``fluxon"$ is the usual electron. Thus so many quasiparticles, electron, $electron^*$, chargeon, fluxon, spinon and holon have been introduced
and hence their bound states must also occur in two reactions,
chargeon+fluxon = electron, chargeon+chargeon, fluxon+fluxon,
chargeon+spinon, chargeon+holon, fluxon+spinon, fluxon+holon, etc.

We wish to discuss if the electromagnetic theory of these new quasiparticles is correctly described and whether the masses have been added correctly.

\noindent{\bf 2.~~Comments.}

{\it (i) Masses }

  Let us write the two reactions.
(a) Electron splits into a $``chargeon"$ and a $``fluxon"$, and
(b) the composite fermion (CF) is made by attaching two fluxons to one electron [2,3].
Therefore, the simple mass equations are,
\beg
m_e = m_\eta + m_\phi - (\Delta_e/c^2)
\eeq
where $m_e$ is the mass of the electron, $m_\eta$ is the mass of the chargeon and $m_\phi$ is that of fluxon. We believe that chargeon and the fluxon are very tightly bound so that there is a binding energy, $\Delta_e$, the mass equivalent of which is $\Delta_e/c^2$. The second equation is the mass of the CF. Since, it is made by attaching two fluxons to one electron, its mass reaction should be,
\beg
m_{CF} = m_e + 2m_\phi - (\Delta_{CF}/c^2)
\eeq
to which the binding energy has beem included. Now we have the burden of measuring the masses,
\beg
(1) m_e,\,\,\, (2)m_\eta,\,\,\, (3) m_\phi,\,\,\, (4)\Delta_e\,\,\, and \,\,\,(5) \Delta_{CF}
\eeq
There are far too many masses to understand the experiments on resistivity in the quantum Hall effect so that adjusting five masses to get the one, experimentally measured mass, 0.4$ m_e$ will not lead to definite values.

{\it(ii) Electromagnetic theory.}

     According to well known Maxwell equations, it is not possible for a charged particle to go round a circle and not produce a field. The chargeon of mass $m_\eta$ is a charged particle but it does not produce a magnetic field. The field is produced by $``fluxon"$ of mass $m_\phi$ but not by $``chargeon"$. Hence,  chargeons do not exist. The $m_{CF}$ requires the existence of $m_\phi$ which according to the earlier cited reaction requires $m_\eta$. Since $``chargeons"$ are excluded by Maxwell equations, the CF are also excluded by the Maxwell equations. So if CF are really existing particles, then there must be a new theory. Such a new theory will have five different masses as discussed above. According to the above discussion, the magnetic vector can be carried by an independent particle of mass, $m_\phi$. So the electric vector will be 
carried by $m_\eta$. This means that {\bf E} and {\bf H} are carried by different quasiparticles. According to Maxwell equations, the time derivative of the electric field determines the {{\it curl}\,\,{\bf H}}
and the time derivative of the magnetic field determines {{\it curl}\,\,{\bf E}}. If {\bf E} and {\bf H} are coupled, the Maxwell equations will be lost, which is contrary to the experimental measurements.

     Therefore, CF are new objects not consistent with Maxwell equations. So the Maxwell equations can be modified and CF can be accepted as a new basic principle but there is no evidence in favour of the variety of masses. When the electron splits into a chargeon and a fluxon, a large binding energy is released but there is no evidence for the release of such energy. Similarly, when CF breaks into an electron and two fluxons, large energy should be released but there is no evidence of such an $``electron \,\,bomb"$ or a $CF ``bomb"$. Indeed, the $``electron \,\,decay"$ is unheard off. If it is possible to attach floxons to electrons,  it must be possible to detach flux from the electron. This gives rise to a ${\bf `` decomposite\,\, fermion"}$ (DF)[4]. The CFs are large objects and can not have the same density as that of the electron. The CF model is therefore internally inconsistent and should be discarded.

\noindent{\bf3.~~ Additional \,\,Comments}.

Ref.5 shows that the experimental identification of CF by Kukushkin et al [6] is not justified. Ref.7 shows the lack of even feature in the data of Pan et al [8]. Ref.9 shows the absence  of ESR in the flux attached field, meaning that flux attachment formula is not correct.
Ref.10 shows that the observed $``opposite\,\,\, spin"$ aspect is not a part of the CF model and Kumada et al [11] have not given the references properly. Ref.12 shows that CF model is inconsistent because it did not consider $``decomposite\,\, fermions"$. Ref.4 points out several inconsistencies in the CF model for example there are far to many parameters. Ref.13 shows that the  spin flip is not a part of the CF model whereas Dujovne et al [14] find them. Ref.[15] shows that {\bf E} and {\bf H} separation is required for the CF model but it does not occur in the data of Pan et al [16]. Ref. 17 shows that $``111"$ is is a fermion and it is not a boson due to antisymmetry and hence the results of Simon et al [18], require to be corrected.

\noindent{\bf4.~~ Conclusions}.

  Ichinose[1] requires that the electron should have structure and the classical electrodynamics should break down. We point out that this type of theory introduces far too many masses to be compared with only one experimental value. The experiments have compared observations with the CF model but in all cases, the identification of the data with the CF is not justified. The stringent requirement of CF model is that flux is attached to the electron. This requirement is not justified and hence the CF model should be discarded.  Ichinose [19-21] has also published several papers on the spin-charge decoupling but these are not applicable to the data on quantum Hall effect. 

\newpage   

\noindent{\bf About the author}: {\it Keshav Shrivastava has obtained Ph.D. degree from the
Indian Institute of Technology and D. Sc. from Calcutta University. 
He is a member of the American Physical Society, Fellow of the Institute of Physics (U.K.)and Fellow of the National Academy of Sciences, India. He has worked in the Harvard University, University of California at Santa Barbara, the University of Houston and the Royal Institute of Technology Stockholm. He has published 170 papers in the last 40 years. He is the author of two books. Shrivastava's paper with Roy Anderson,
published in J. Chem. Phys. 48, 4599(1967) was found to be useful by M. C. R. Symons,F. R. S. The letter with K. W. H. Stevens, J. Phys. C 3, L 64 (1970) was useful to D. I. Bolef, the paper with Vincent Jaccarino,Phys. Rev. B13, 299(1976) was useful to J. B. Goodenough, F. R. S. His paper, J. Phys. C 20, L789 (1987) on the microwave absorption, was useful to K. Alex M\"uller. His paper published in the Proc. Roy. Soc. A419, 287- 303(1988) was communicated by B. Bleaney, F.R.S. He discovered the flux quantized energy levels in superconductors and the correct theory of 1/3 charge in quantum Hall effect}.

The correct theory of quantum Hall effect is given in ref.22.
\vskip1.25cm

\noindent{\bf5.~~References}
\begin{enumerate}
\item Ikuo Ichinose and Tetsuo Matsui, cond-mat/0210142.
\item J. K. Jain, Phys. Rev. Lett. {\bf 63}, 199 (1989).
\item K. Park and J. K. Jain, {\bf 80}, 437 (1998)
\item K. N. Shrivastava, cond-mat/0210320.
\item K. N. Shrivastava, cond-mat/0202459.
\item I. V. Kukushkin et al, Nature {\bf 413}, 409 (2002).
\item K. N. Shrivastava, cond-mat/0204627.
\item W. Pan et al, Phys. Rev. Lett. {\bf 88}, 176802(2002).
\item K. N. Shrivastava, cond-mat/0207391.
\item K. N. Shrivastava, cond-mat/0209057
\item N. Kumada, et al, Phys. Rev. Lett. {\bf 89}, 116802(2002).
\item K. N. Shrivastava, cond-mat/0209666
\item K. N. Shrivastava, cond-mat/0211223.
\item Dujovne, et al, cond-mat/0211022.
\item K. N. Shrivastava, cond-mat/0301380.
\item W. Pan et al, Phys. Rev. Lett. {\bf 90}, 016801(2003)
\item K. N. Shrivastava, cond-mat/0302009.
\item S. H. Simon et al, cond-mat/0301203.
\item I. Ichinose, T. Matsui and M. Onoda, Phys. Rev. B {\bf 64},104516 (2001).
\item I. Ichinose and T. Matsui, Phys. Rev. Lett. {\bf 86}, 942 (C)(2001).
\item I. Ichinose and M. Matsui, Phys. Rev. B {\bf 51}, 11860 (1995).
\item K.N. Shrivastava, Introduction to quantum Hall effect,\\ 
      Nova Science Pub. Inc., N. Y. (2002).
\end{enumerate}
\vskip0.1cm
Note: Ref. 22 is available from:\\
 Nova Science Publishers, Inc.,\\
400 Oser Avenue, Suite 1600,\\
 Hauppauge, N. Y.. 11788-3619,\\
Tel.(631)-231-7269, Fax: (631)-231-8175,\\
 ISBN 1-59033-419-1 US$\$69$.\\
E-mail: novascience@Earthlink.net

\vskip5.5cm
\end{document}